\documentstyle[12pt]{article}

\newcounter{saveeqn}
\newcommand{\alpheqn}{\setcounter{saveeqn}{\value{equation}}%
\stepcounter{saveeqn}\setcounter{equation}{0}%
\renewcommand{\theequation}{\mbox{A\arabic{equation}}}}
\newcommand{\reseteqn}{\setcounter{equation}{\value{saveeqn}}%
\renewcommand{\theequation}{\arabic{equation}}}

\begin{document}
\section*{The Low-lying Excitations of Polydiacetylene}
 A. Race $^{1}$, W. Barford $^{1}$ and R. J. Bursill $^{2}$

$^{1}${\it Department of Physics and Astronomy, The University of
Sheffield, Sheffield, S3 7RH, United Kingdom.}

 $^{2}${\it School of
Physics, The University of New South Wales, Sydney, NSW 2052,
Australia.}

\bigskip
\begin{abstract}
The Pariser-Parr-Pople Hamiltonian is used to calculate and
identify the nature of the low-lying vertical transition energies
of polydiacetylene. The model is solved using the density matrix
renormalisation group method for a fixed acetylenic geometry for
chains of up to 102 atoms. The non-linear optical properties of
polydiacetylene are considered, which are determined by the
third-order susceptibility. The experimental 1{\it B}$_{u}$ data
of Giesa and Schultz are used as the geometric model for the
calculation. For short chains, the calculated {\it E}(1{\it
B}$_{u}$) agrees with the experimental value, within solvation
effects ($\sim $ 0.3 eV). The charge gap is used to characterise
bound and unbound states. The {\it nB}$_{u}$ is above the charge
gap and hence a continuum state; the 1{\it B}$_{u}$, 2{\it
A}$_{g}$ and {\it mA}$_{g}$ are not and hence are bound excitons.
For large chain lengths, the {\it nB}$_{u}$ tends towards the
charge gap as expected, strongly suggesting that the {\it
nB}$_{u}$ is the conduction band edge. The conduction band edge
for PDA is agreed in the literature to be $\sim $ 3.0 eV.
Accounting for the strong polarisation effects of the medium and
polaron formation gives our calculated {\it E$_{\infty} $}({\it
nB}$_{u}$) $\sim $ 3.6 eV, with an exciton binding energy of
$\sim $ 1.0 eV.{\it} The 2{\it A}$_{g}$ state is found to be
above the 1{\it B}$_{u}$, which does not agree with relaxed
transition experimental data. However, this could be resolved by
including explicit lattice relaxation in the
Pariser-Parr-Pople-Peierls model. Particle-hole separation data
further suggest that the 1{\it B}$_{u}$, 2{\it A}$_{g}$ and {\it
mA}$_{g}$ are bound excitons, and that the {\it nB}$_{u}$ is an
unbound exciton.
\end{abstract}

\subsection{INTRODUCTION}
\label{sub.intro}

\bigskip

Polymers, and other molecular materials, that exhibit non-linear
optical properties and electroluminescence have attracted much
interest amongst theorists and experimentalists, owing to their
possible use in organic technology~\cite{leng}--\cite{barf1}.
Poly(diacetylene)s (PDAs) with the general formula shown in Fig.
1. (a) are the $\pi $-conjugated organic polymers studied here:
they exhibit non-linear optical properties, near-perfect crystal
structure and doping-dependent transport properties. These
properties make them ideally suited for use in the theorist's
calculation. A full characterisation of PDA's electronic
properties would mean a better understanding of $\pi $-conjugated
polymers in general. In PDA with a very low polymer content ({\it
x}$_{p}< 10^{-3}$ in weight), interchain interaction is minimal
and thus it can be considered an ideal one-dimensional model
system.

There have been many calculations performed on PDA with varying
degrees of success. Parry conducted a SCF calculation and found
an energy gap that was too low \cite{parry}. Boudreaux's SCF-{\it
X$\alpha $} calculation gives an energy gap of 2.21 eV, which
agrees well with the long-chain experimental extrapolations for
the energy gap of 2.11 eV and 2.25 eV (yellow phase)
\cite{gross}--\cite{imp.giesa}. A further SCF-{\it X$\alpha $}
calculation by Boudreaux \cite{wine1} accounts for charge density
waves by the same geometric H\"{u}ckel model of Whangbo
\cite{whangbo}. The {\it ab initio} crystal orbital calculations
of Kertesz give energy gaps that are too high
\cite{kertesz1}--\cite{kertesz2}. Predicted bond lengths show
impressive agreement with x-ray data in Karpfen's {\it ab initio}
SCF calculation \cite{karpfen}. He performed a total energy
minimisation to calculate bond lengths and shows that the energy
difference between butatriene and acetylene structure per unit
cell is 0.52 eV. This agrees with previous calculations and
experiment \cite{whangbo,sixl}. A similar calculation yielding
similar results was performed by Br\'{e}das \cite{bredas}.

Wilson, Cojan, Cade and Movaghar \cite{wilson}--\cite{cade}
performed H\"{u}ckel band structure calculations for the
infinitely long PDA chain. Cajon and Wilson did not include bond
alteration in their calculation, and thus modelled a compound
with one distinct chemical bond. The energy gap was found to be
too low. Cade and Movaghar used two degrees of freedom in their
calculation as a bond alteration parameter.

Extended H\"{u}ckel calculations have been performed by Parry and
Whangbo \cite{whangbo,parry2}. The work of Parry is restricted to
one distinct chemical structure and, as with other
parameterisations, yields a value for the energy gap that is too
low. The model employed by Wangbo has a hopping term of the form:
\begin{equation}
t=t_0(1\pm\delta)
\end{equation}
with 0 $< \delta <$ 1. This produces results that are, again,
quantitatively too low; however, it does generate an energy
difference of 0.48 eV between the butatriene and acetylene
structures, in excellent agreement with other calculations and
experiment \cite{karpfen}--\cite{sixl}. In summary, therefore, all
these models are seen to give good qualitative predictions for
bond alteration, but the data are seen to be consistently
red-shifted owing to their single particle picture.

The importance of screening in PDA is still considered
controversial to many authors, because it determines the role of
electron-electron correlations. A simple way to account for
screening is to adjust the single chain electron-electron
interaction potential {\it V}({\it r}), a method suggested by
many authors. For example, Abe lowered the Hubbard parameter {\it
U} from 11 to 5 eV and used a large long-range dielectric
constant in a Ohno potential \cite{abe}--\cite{moore}. However,
this again causes red-shifted energy gaps and results in
unphysical atomic cohesion energies. We emphasise that there is
clear and indisputable evidence that electron-electron
interactions are of paramount importance in the electronic
structure calculations of PDA. First, in polymeric crystal form,
the linear absorption spectrum is see to be symmetric and peaks
at $\sim $ 1.8-1.9 eV. However, the onset of photoconduction
occurs not at this value but at $\sim $ 2.3-3.0 eV (corresponding
to uncorrelated $\pi $-$\pi $* transitions) \cite{yee}. In
addition, comparison of the linear absorption with the EA
spectrum shows an EA feature at $ \sim $ 2.0 eV corresponding to
the first derivative of the absorption: this is attributed to the
Stark shift of the exciton. There is also an oscillating
Franz-Keldysh band-edge structure at $ \sim $ 2.5-3.0 eV, with a
lineshape that deviates from the first derivative of the
absorption. The oscillations are due to interband transitions and
coincide with the onset of photoconduction. These data
independently suggest that the exciton binding energy is $\sim $
0.5 eV \cite{sebastian}--\cite{hasegawa}. We note that these
values show a great deal of variability owing to the variety of
phases and the disorder present in PDA. Hence, despite the
apparent success of the Su-Schrieffer-Heeger model
\cite{ssh1}--\cite{ssh2} neutral excitons with large binding
energies are the low-lying excitations in PDA \cite{yee},
\cite{guo.muzumdar}--\cite{moses}. This is borne out by the
Pariser-Parr-Pople calculation presented in this work, which
incorporates electron-electron interactions.

\bigskip
\subsection{MODELLING POLYDIACETYLENE}
\label{sub.model}

\bigskip
\subsubsection{Electronic Structure}
\label{sub.sub.elect}

\bigskip

The atomic orbital model used in the calculation is described as follows.
Carbon has the atomic configuration 1{\it s}$^{2}$2{\it s}$^{2}$2{\it
p}$^{2}$. Since carbon has four valence electrons, there is an alternating
structure of two {\it sp}$^{2}$ and two {\it sp} hybridised carbon atoms
that can account for the idealised PDA chemical composition. (See Fig. 1.)

Since PDA is spatially centrosymmetric, and thus shows {\it C}$_{2}$
symmetry, the wavefunctions possess mirror plane and centro-inversion
symmetries. The group notation for mirror plane symmetries are {\it A} for
the symmetric and {\it B} for the antisymmetric case. Inversion symmetries
are labelled {\it g} for symmetric and {\it u} for antisymmetric. The ground
state is therefore labelled 1{\it A}$_{g}$ and the first optically active
dipole from the ground state has to be the 1{\it B}$_{u}$ state. The
wavefunctions are either even ({\it A}$_{g}$) or odd ({\it B}$_{u}$) under
an inversion.

\bigskip
\subsubsection{Geometric Structure}
\label{sub.sub.geom}

\bigskip

Although there is a myriad of literature on the optical
properties of PDA, there is still little on their structural and
conformational properties, which prevents extensive theoretical
investigation. The work of Giesa and Schultz \cite{imp.giesa}
following that of Wudl and Bitler \cite{wudl.bitler} describes the
synthesis and thorough characterisation of the series of
alternating all-{\it trans}-polyenynes without substituents at
the vinylic bonds that are studied in this paper. 

To maintain consistency with the literature, the unsubstituted model
compounds are named as follows:

\begin{equation}
C_{n}[{\it N}]
\end{equation}

\bigskip

\noindent where the number of unsaturated carbon atoms {\it n}
specifies each compound, and the number of formal monomer units
{\it N} are added in the square brackets for clarity. The
relation between {\it n} and {\it N} is,

\begin{equation}
n=2(2{\it N}+1). \label{N}
\end{equation}

\bigskip
These structures are shown in Fig. 2. In contrast with the
simplest conjugated polymer represented by {\it
trans}-polyacetylene, the carbon backbone of PDA contains two,
additional, localized $\pi $ electrons in every unit cell. Thus,
there are two possibilities of bond alternation in this system
that lead to non-equivalent structures with non-degenerate ground
states, as shown in Fig. 1. (b){\bf} and (c). Work on the molecular
geometries of C$_{14}$[3] and C$_{22}$[5] by X-ray structure
analysis has determined bond lengths and angles that represent a
typical {\it acetylene} structure (see TABLE I.) and no evidence
for a {\it butatriene} form is found. In addition, Giesa shows 
that substitution does not alter the polymer structure significantly. Increasing the
size of the polymer does elongate double and triple bonds, while
reducing the single bond; however, this is not pronounced. Hence,
overall, these findings justify the use of the model acetylene
structure employed in this paper, because an increase in
conjugation length does not show a substantial effect on bond
lengths and angles. Hence, owing to PDA's simple electronic and
geometric ground-state structure, the Pariser-Parr-Pople
Hamiltonian can easily describe it.

\bigskip

\subsection{THE PARISER-PARR-POPLE HAMILTONIAN}
\label{sub.PPP}

The Pariser-Parr-Pople Hamiltonian is:
\begin{eqnarray}
\hat{H}_{PPP} & = & -\sum_{<ij>\sigma}t_{ij}(c_{i\sigma}^\dagger
c_{j\sigma} + c_{j\sigma}^\dagger c_{i\sigma})+
U\sum_{i}(n_{i\uparrow} - \frac{1}{2})(n_{i} - \frac{1}{2})
\nonumber \\
& + & \sum_{i \neq j \atop \sigma \sigma '}V_{ij}(n_{i \sigma}
-1)(n_{j \sigma '} -1)
\label{PPP}
\end{eqnarray}

\bigskip
where the operator $c_{i\sigma}^\dagger (c_{i\sigma})$ creates
(annihilates) a 2{\it p}$_{z}$ electron of spin $\sigma $ at site
{\it i} (the {\it i}th carbon atom), {\it t}$_{i,j}${\it} ($ >
$0) is the transfer integral between the nearest neighbour atomic
orbitalsl. The first term in the Hamiltonian allows the electrons
to \lq\lq hop\rq\rq \ from one atom to another, and represents the
kinetic energy gained from delocalising an electron from its
atomic site. $n_{i \sigma} = c_{i \sigma} c_{i \sigma}^{
\dagger}$ is the number density operator, and $<{\it ij}>$
denotes nearest neighbours. {\it U} is the Pariser-Parr-Pople
Coulomb repulsion between two electrons occupying the same 2{\it
p}$_{z}$ orbital and {\it U} = 10.06 eV
\cite{bursill.castleton.barf}. {\it V}$_{ij}$ is the long-range
Coulomb repulsion and, in this work, is the Ohno function,
$V_{ij}=U/ \sqrt{1+ \beta r_{ij}^{2}}$ where $\beta ${\it} =
({\it U/}14.397)$^{2}$ and {\it r}$_{ij}$ is the interatomic
distances measured in {\AA} \cite{ohno}.

The general expression for the resonance integrals
is given by:
\begin{equation}
t_{ij}=t_{0} + \alpha (r_{ij}-r_{0}) \label{linearhop}
\end{equation}
where $\alpha ${\it} is the coupling constant of electron-phonon
interactions, {\it r}$_{ij}$ is the length of the bond between
carbon atoms {\it i} and {\it j}, and {\it r}$_{0}$ is the
average bond length of the reference system. 

The parameters in expression (\ref{PPP}) are derived using
polyacetylene as a reference system. First, it is pertinent to
consider the gradient of the resonance integral with respect to
bond length. It has been shown that the difference $ \delta t $
of the resonance integral is 0.2162 eV and the corresponding
average change in bond lengths $ \delta l$ is 0.0471 Å
\cite{bursill.barf1}. Thus, the electron-phonon coupling constant
is:
\begin{equation}
\alpha = \frac{ \delta t}{ \delta l} = \frac{0.2162}{0.0471} \
{\rm eV}
\end{equation}
A full list of Hamiltonian parameters derived from polyacetylene
used in this work is shown in TABLE II., and a schematic of the
unit cell is found in Fig. 3.

The different resonance integrals are determined by the specific
structures of the chain with lattice constant 4{\it a}.
Perturbations of the equidistant carbon chain lead to the energy
gaps {\it E}$_{g} $ and {\it E}$_{g}^{'}$. The ground state
corresponds to a fully filled valence band and unoccupied
conduction band. The 1{\it B}$_{u}$ is reached from the ground
state by exciting one electron from the highest occupied
molecular orbital (HOMO) to the lowest unoccupied molecular
orbital (LUMO) (see Fig. 4.).

Although this band picture is useful in identifying important
states, it has been shown that interactions {\it do} play an
important part in the physics of PDA, as mentioned earlier. By
including the on-site and long range terms explicitly in the
Hamiltonian (\ref{PPP}) electron-electron interactions are taken
into account, giving an accurate treatment of the electronic
structure of PDA. Turning on the interactions, {\it U} and {\it
V}$_{ij}$, changes the relative position of electronic states and
can even cause the crossing of states. This non-interacting
picture is therefore modified and an excitonic picture
formulated. An electron and hole can bind together by their
Coulomb field to produce an exciton, or bound electron-hole pair.
Since the conduction band signifies the non-interacting limit,
excitons are energetically situated below the conduction band.

\bigskip

\subsection{RESULTS AND DISCUSSION}
\label{sub.RandD}

\bigskip

\subsubsection{Essential States}
\label{sub.sub.essential}

\bigskip

Non-linear optical experimental data can be used to characterise
the low-lying states of $\pi $-conjugated polymers. The relative
ordering of these states can elucidate much of the physics of
such systems \cite{guo.muzumdar},
\cite{guo.dandan.mazum}--\cite{guo.dandan.mazum2}. Non-linear
optical processes in polymers with {\it C}$_{2}$ symmetry are
determined by the third-order susceptibility, $\chi ^{3}
(-\omega_{1}-\omega_{2}-\omega_{3};\omega_{1},\omega_{2},\omega_{3})$,
which is calculated by the sum over all available states. This
makes calculating $\chi ^{(3)}$ in principle difficult, owing to
the large number of paths included. However, it is found that
there are certain {\it essential states} that are important when
considering the $\chi ^{(3)}$ of $\pi $-conjugated polymers, as
only the essential states make a significant contribution to it.
The four essential states are the ground state, the first
odd-parity exciton state, 1{\it B}$_{u}$, the charge transfer
state, {\it mA}$_{g}$, and the {\it nB}$_{u}$ (see Fig. 5.).
Another important state is the 2{\it A}$_{g}$: this is a
two-photon state, whose position relative to the {\it 1B}$_{u}$
determines whether a polymer exhibits photoluminescence. The
importance of these states is the result of the strong and
dominant dipole couplings amongst them in
\begin{equation}
\mu_{if}=\langle f| \hat{\mu} |i \rangle
\end{equation}
resulting in their substantial contribution to non-linear optical
spectroscopy. Hence, an important first step in identifying the
non-linear properties of PDA is calculating the dipole moments of
transitions between states. As the DMRG method is a robust and
accurate method of finding dipole moments the essential states
can be found. These are given in TABLE III. However, they are
only calculated for polymers of up to 26 sites. This is because
the important optical states become interlaced with other states
of a spin-density-wave character (i.e., those related to the 2{\it
A}$_{g}$). Since we can only target approximately 10 states in
each symmetry sector, the {\it mA}$_{g}$ and the {\it nB}$_{u}$
states soon become impossible to track.

An analysis of the essential states and the 2{\it A}$_{g}$ forms
the remainder of this discussion. They are paramount when
considering the physics of $\pi $-conjugated polymers, as
mentioned above, and are compared with experimental data. In
Section \ref{sub.RandD} (\ref{sub.sub.vertical}) the vertical
excitation energies of the essential states and the 2{\it
A}$_{g}$ are analysed. Section \ref{sub.RandD}
(\ref{sub.sub.particle}) contains particle-hole separation data,
which is used to confirm predictions made in Section
\ref{sub.RandD} (\ref{sub.sub.vertical}) and help identify the
nature of the 1{\it B}$_{u}$, 2{\it A}$_{g}$, {\it nB}$_{u}$ and
{\it mA}$_{g}$ states.

\bigskip
\subsubsection{Vertical Excitation Energies}
\label{sub.sub.vertical}

\bigskip

The DMRG method, used by Barford and Bursill \cite{barf1},
following the work of White \cite{white1}, was implemented on the
model system to find the vertical low-lying energy eigenvalues of
PDA. The low-lying excitations are bound particle-hole pairs,
which behave as composite particles. These particles delocalise
along the polymer backbone as effective single particles, and
thus their dispersion should be characteristic of single
particles. In PDA the excitons effectively tunnel between double
and triple bond dimers. In an effective particle model, this
would be modelled as a linear chain, with two `sites' per unit
cell. The dispersion should scale as $1/{\it n}^{2}$, as these
are states {\it within} the exciton band. In contrast, the band
gap is associated with free, single particle transitions across
the HOMO-LUMO gap. An examination of the dispersion relation of
Lennard-Jones \cite{lennard} indicates that this energy should
scale as $1/ {\it n}$ in the asymptotic limit. However, neither a
solely 1/{\it n} nor 1/{\it n}$^{2}$ fit for the free
particle-hole and exciton dispersions, respectively, is accurate
for small chains, owing to higher-order corrections, and the fit
required cannot be used to characterise states. In this work,
therefore, a polynomial fit is preferred for extrapolating the
data.

The calculated vertical excitation energies for the 1{\it
B}$_{u}$, 2{\it A}$_{g}$, {\it mA}$_{g}$, {\it nB}$_{u}$ and
charge gap, $\Delta $, are plotted in Fig. 6. against 1/{\it n}.
In addition, the polynomial extrapolations for the long-chain
limit are shown in TABLE IV. for some of the states. The charge
gap is a useful criterion for characterising excitonic and unbound
states, and is given as follows,
\begin{equation}
\Delta  = {\it E}({\it n}\,+ 1)+{\it E}({\it n}-1)-2{\it E}({\it
n}\, )
\end{equation}
Here {\it E}({\it n}) is the ground-state energy of the {\it n}
electron system. The charge gap signifies the lowest energy
excitation of an electron from the valence to conduction band.
Thus states above $\Delta $ are unbound excitons; those below it
are bound. Of the states plotted in Fig. 6. only the {\it
nB}$_{u}$ is positioned above the charge gap, suggesting that the
{\it nB}$_{u}$ is a continuum state. Conversely, the 1{\it
B}$_{u}$, 2{\it A}$_{g}$ and {\it mA}$_{g}$, are below it and are
thus bound excitons. This is shown to be true in Section
\ref{sub.RandD} (\ref{sub.sub.particle}) by examining the
particle-hole separation of these states.

The experimental data obtained by Giesa \cite{imp.giesa} are also
plotted in Fig. 6. A polynomial extrapolation of the experimental
data yields {\it E$_{\infty} $} $ \sim $ 2.5 eV; however, this is
considered unreliable as there are so few points used in the fit.
For short chains the calculated and experimental 1{\it B}$_{u}$
energies are remarkably within solvation (polarisation) effects
of the experimental medium ($\sim $ 0.3 eV). This solvation value
is derived from the work of Yaron, who predicts 0.3 eV for the
1{\it B}$_{u} $ \cite{moore}. In addition, Barford and Bursill's
calculation on polyenes \cite{bursill.barf1} found {\it E}\,(1{\it
B}$_{u}$) to be 0.3 eV above the experimental value. Their recent
calculations on PPP and PPV yield similar results
\cite{unpub.barf}. Our calculated {\it E}$_{\infty}$(1{\it
B}$_{u}$) for PDA is $ \sim $ 3.0 eV, and, hence, correcting for
solvation effects gives our calculated {\it E$_{\infty} $}(1{\it
B}$_{u}$) $ \sim $ 2.7 eV.

The energy of the {\it nB}$_{u}$ is seen to tend to that of the
charge gap for {\it n}\ $\to\infty$, as seen in Fig. 6. Hence, the
extrapolated long chain {\it nB}$_{u}$ energy is found to be $
\sim $ 5.7 eV, and this state is predicted to be the conduction
band threshold. However, the onset of photoconduction (continuum
limit) is $ \sim $ 3.0 eV. The large polarisation effects of the
surrounding medium can correct for this: these are estimated by
Yaron to be $ \sim $ 1.5 eV \cite{moore}. In addition, polaron
formation means a further $ \sim $ 2 $ \times $ 0.3 eV can be
subtracted from 5.7 eV \cite{bursill.barf1}. Including all these
effects gives {\it E$_{\infty}$}({\it nB}$_{u}$) $ \sim $ 3.6 eV,
which is in reasonable agreement with the experimental conduction
band threshold quoted earlier. Therefore, the exciton binding
energy is {\it E}$_{b}$ = {\it E$_{\infty}$}({\it nB}$_{u}$) $-$
{\it E$_{\infty} $}(1{\it B}$_{u}$), and from our calculation is
found to be $3.6-2.7 \sim 0.9$ eV. This is in reasonable
agreement with binding energies found in the literature
\cite{sebastian}--\cite{hasegawa}.

Having identified the important non-linear optical states in PDA
from the dipole moments, and calculated the corresponding
excitation energies, an energy level diagram can be drawn. This
is shown for 26 atoms in Fig.7. As expected there is a large
1{\it A}$_{g}\to \; $1${\it B}_{u}$ dipole moment, corresponding
to the one-photon absorption, while the band threshold ({\it
nB}$_{u}$) has a weak coupling to the ground state. It is seen
that the first excited {\it A}$_{g} $ state is above the 1{\it
B}$_{u}$, which would suggest that PDA shows photoluminescence.
(There is a direct 1{\it A}$_{g}\to $ $1${\it B}$_{u}$
transition.) However, experiments have shown that PDA is not
electro-luminescent: measurements of the frequency dependence of
third harmonic generation in Langmuir-Blodgett films of PDA
indicate that an {\it A}$_{g}$ singlet state is $\sim $ 0.11 eV
below the lowest energy {\it B}$_{u}$ state. In addition, the
recent work of Kohler and Schilke \cite{kohler.schilke} has shown
an {\it A}$_{g}$ symmetry state $\sim $ 0.21 eV lower than the
lowest {\it B}$_{u}$ symmetry state. Our model calculates
vertical transitions only and is seen to agree with many
theoretical vertical-energy calculations of PDA. Singlet exciton
relaxation studies on isolated polydiacetylene chains by
subpicosecond pump-probe experiments have suggested that
nonradiative singlet {\it B}$_{u} $ exciton relaxation involves
two {\it A}$_{g}$ states in series \cite{kraabel.joffre}. As it
is well known that electron-phonon coupling is strong in PDA, we
propose that lattice relaxation effects produce these {\it
A}$_{g}$ states. It is hoped that this will be resolved by
studies of lattice relaxation explicitly within the
Pariser-Parr-Pople-Peierls Hamiltonian, which are now in
progress. Similar studies by Barford on polyenes have found that
the relaxation energies of the 1{\it B}$_{u}$ and 2{\it A}$_{g} $
can be reduced by as much as 0.3 and 1.0 eV, respectively
\cite{barf.burs.lav}--\cite{lav.barf}. Thus, including lattice
relaxation in the Pariser-Parr-Pople-Peierls model is expected to
bring the 2{\it A}$_{g}$ state below the 1{\it B}$_{u}$ in
agreement with experiment.

\bigskip
\subsubsection{Particle-hole Separation}
\label{sub.sub.particle}

\bigskip

In order to help characterise the low-lying {\it A}$_{g}$ and
{\it B}$_{u}$ states the particle-hole separation is calculated
and gives an indication of the spatial extent of a given state:
an explanation of how it is calculated is given in the Appendix.
Suffice it here to say that we use the one-particle singlet
excitation correlation function, which directly relates a hole in
the valence band to an electron in the conduction band. In
essence, if a particular state's particle-hole separation is seen
to increase with system size, the electron and hole are unbound.
However, if this quantity reaches a maximum, the Coulomb
attraction between the electron and hole is strong enough to bind
them together. In Fig. 8. the particle-hole separations are given
as a function of system size in units of the average C-C bond
length for the 1{\it B}$_{u}$, 2{\it A}$_{g}$ and {\it mA}$_{g}$
states. It is clear that the 1{\it B}$_{u}$, 2{\it A}$_{g}$ and
{\it mA}$_{g}$ states are all bound excitons because the
electron-hole separation reaches a maximum value with increasing
chain length. These signify the composite particles mentioned in
Section \ref{sub.RandD} (\ref{sub.sub.vertical}). However, the
particle-hole separation of the {\it nB}$_{u}$ state increases
linearly with system size: this suggests that the {\it nB}$_{u}$
is either very weakly bound or unbound. We conclude therefore
that for PDA molecules of up to 26 carbon atoms the electron-hole
continuum (i.e., the band edge) is expected to be the {\it
nB}$_{u}$, further strengthening the arguments of Section
\ref{sub.RandD} (\ref{sub.sub.vertical}).

\bigskip

\subsection{CONCLUSIONS}

The DMRG calculation of a suitably parameterised
Pariser-Parr-Pople Hamiltonian for a {\it rigid geometry} can be
used to describe the electronic structure of short or long chain
polydiacetylene polymers. The role of electron-electron
interactions is fundamental in the low-lying excitations, and
this is borne out in our calculation. We have found and
characterised the essential states: these are the 1{\it A}$_{g}$,
1{\it B}$_{u}$, {\it mA}$_{g}$ and {\it nB}$_{u}$. Our 1{\it
B}$_{u}$ excitation energies are within a few tenths of an eV of
the experimental ones: for short chains, the calculated {\it E}
\,(1{\it B}$_{u}$) agrees with the experimental value, within
solvation effects ($\sim $ 0.3 eV). The {\it nB}$_{u}$ is found
to be above the charge gap, and hence it is continuum state; the
1{\it B}$_{u}$, 2{\it A}$_{g}$ and {\it mA}$_{g}$ are not, and
hence are bound excitons. For large chain lengths the {\it
nB}$_{u}$ tends towards the charge gap as expected, strongly
suggesting that the {\it nB}$_{u}$ is the conduction band edge.
We found our calculated conduction band edge to agree reasonably
with the experimental value of $\sim $ 3.0 eV quoted in the
literature. Accounting for the strong polarisation effects of the
medium and polaron formation gives our calculated {\it E$_{\infty}
$}({\it nB}$_{u}$) $\sim $ 3.6 eV, with an exciton binding energy
of $\sim $ 1.0 eV.{\it} Our 2{\it A}$_{g}$ is calculated to be
above the 1{\it B}$_{u}$, which does not agree with
relaxed-transition experiments: it is hoped that electron-lattice
relaxation will adjust the position of the 2{\it A}$_{g}$ to
correct for this. Particle-hole separation data were used to help
characterise the low-lying excitation in PDA. These data further
suggest that the 1{\it B}$_{u}$, 2{\it A}$_{g}$ and {\it
mA}$_{g}$ are bound excitons, and that, conversely, the {\it
nB}$_{u}$ is an unbound exciton.

This work is further evidence that the essential states mechanism is
adequate for describing the electronic properties of model $\pi $-conjugated
polymers. Future work will include, as already discussed, the incorporation
of electron-lattice relaxation, and a more thorough treatment of the triple
bond using a ZINDO Hamiltonian.

\subsection*{ ACKNOWLEDGEMENTS}

We thank David Yaron for discussions. One of the authors (A. R.) is
supported by the EPSRC (UK). The calculations were performed on the DEC8400
at the Rutherford Appleton Laboratory.

\subsection*{APPENDIX}
\alpheqn To measure the particle-hole separation we use the
particle-hole correlation function, introduced in Refs. [48] and
[49]. Underlying this approach is the assumption that a
particle-hole pair corresponds to the promotion of an electron
from the valence (or HOMO) band to the conduction (or LUMO) band.

Without loss of generality, the atomic basis can be transformed
to a local molecular orbital basis. The local molecular orbitals
correspond to the bonding ($|1\rangle$) and anti-bonding
$|2\rangle$) combinations of the atomic orbitals on each dimer,
where a dimer is a double or triple bond. The local molecular
orbitals delocalise via the single bonds. Then, the creation
operator,
\begin{equation}
\nonumber
 S^{\dagger}_{ij}=\frac{1}{\sqrt{2}} \left(
a^{\dagger}_{i2\uparrow}a_{j1\uparrow}+a^{\dagger}_{i2\downarrow}a_{ji\downarrow}
\right) \nonumber
\end{equation}
promotes an electron from $|1\rangle$)on dimer {\it j.} to
$|2\rangle$) on dimer {\it i}. However, since the local molecular
orbitals are not exact Bloch transforms of the band molecular
orbital\footnote{ The exact Bloch transforms of the band
molecular orbitals are Wannier molecular orbitals. The Wannier
orbitals of the valence band are predominately $|1\rangle$), with
a small admixture of $|2\rangle$) from neighbouring dimers.\par}
it is also necessary to include the hermitian conjugate of {\it
S}$_{ij}${\it .}

We now define the exciton correlation function with respect to the ground
state, as:
\begin{equation}
C^{s}_{ij}(|n\rangle)= \langle n|
(S^{\dagger}_{ij}+S_{ij})|1^{1}A^{+}_{g} \rangle
\end{equation}
and the mean square of the particle-hole separation is:
\begin{equation}
\langle (i-j)^{2} \rangle = \frac{\sum _{ij} (i-j)^{2}
(C^{s}_{ij})^{2}}{\sum _{ij} (C^{s}_{s})^{2}}
\end{equation}
In practice, we do not consider all combinations of {\it i} and
{\it j,} but restrict ourselves to {\it i} and {\it j}
symmetrically spaced around the middle dimer. \reseteqn

\pagebreak   

\pagebreak

\subsection*{FIGURES}

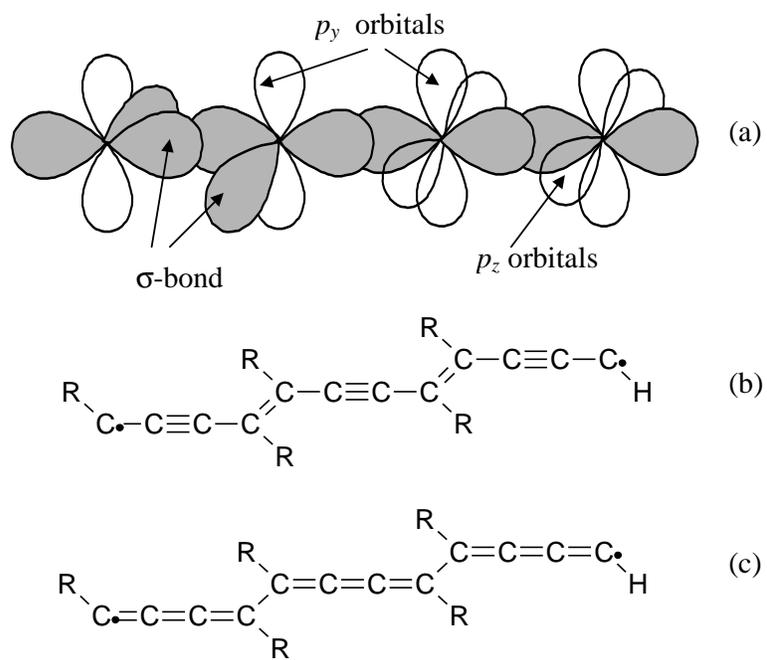
FIG. 1. Chemical structure of a diacetylene trimer molecule. (a)
Schematic representation of the orbitals giving rise to the
different structures. The shaded orbitals give rise to $\sigma
$-bonds, while the unshaded orbitals contain $\pi $-electrons in
2{\it p} orbitals. The 2{\it p}$_{z}$ orbitals all overlap along
the length. The molecular plane is in the {\it x}-{\it y }plane.
(b) Acetlyene structure. (c) Butatriene structure.

\bigskip

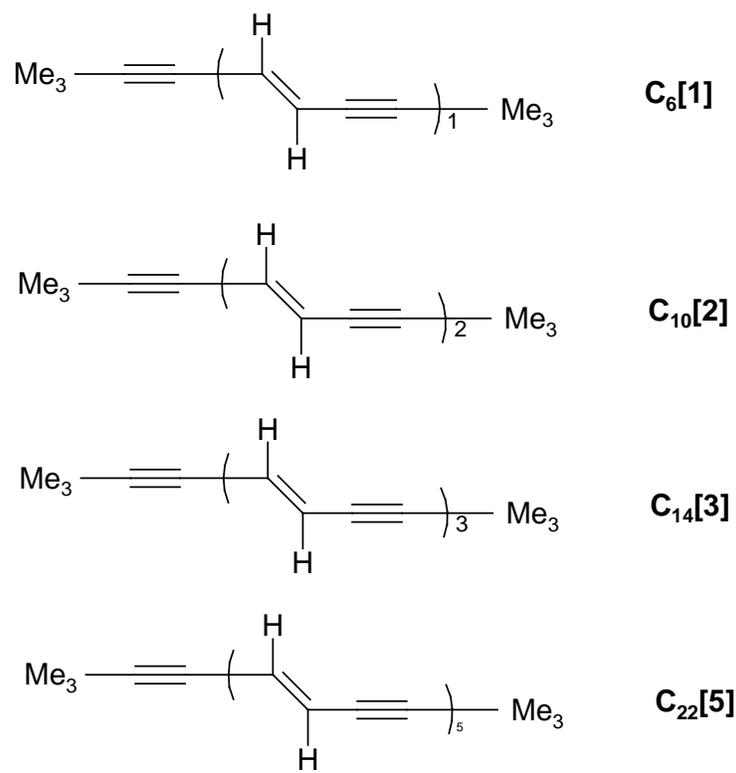
FIG. 2. The idealised chemical structure of the series of PDAs
used in this work. Me$_{3}$ represents three methyl structures.
C$_{n}$[{\it N}] unambiguously represents the structures, where
{\it n} is the number of carbon atoms and {\it N} is the number
of monomer units.

\bigskip

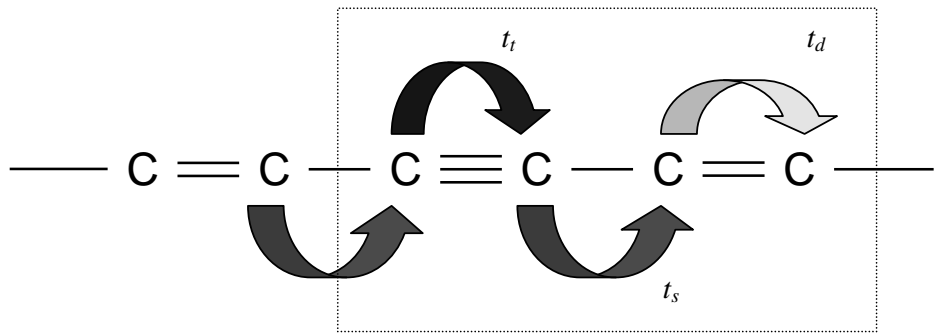
FIG. 3.{\bf}  Resonance integrals used for PDA with bond
alteration in the {\it l}th unit cell. {\it t}$_{s} $is the
resonance integral for the single, {\it t}$_{t}${\it} for the
triple, and {\it t}$_{d}${\it} is for the double bond. The
periodicity of polydiacetylene is given by 4{\it a}, where {\it
a} is the undimerized bond length.

\bigskip

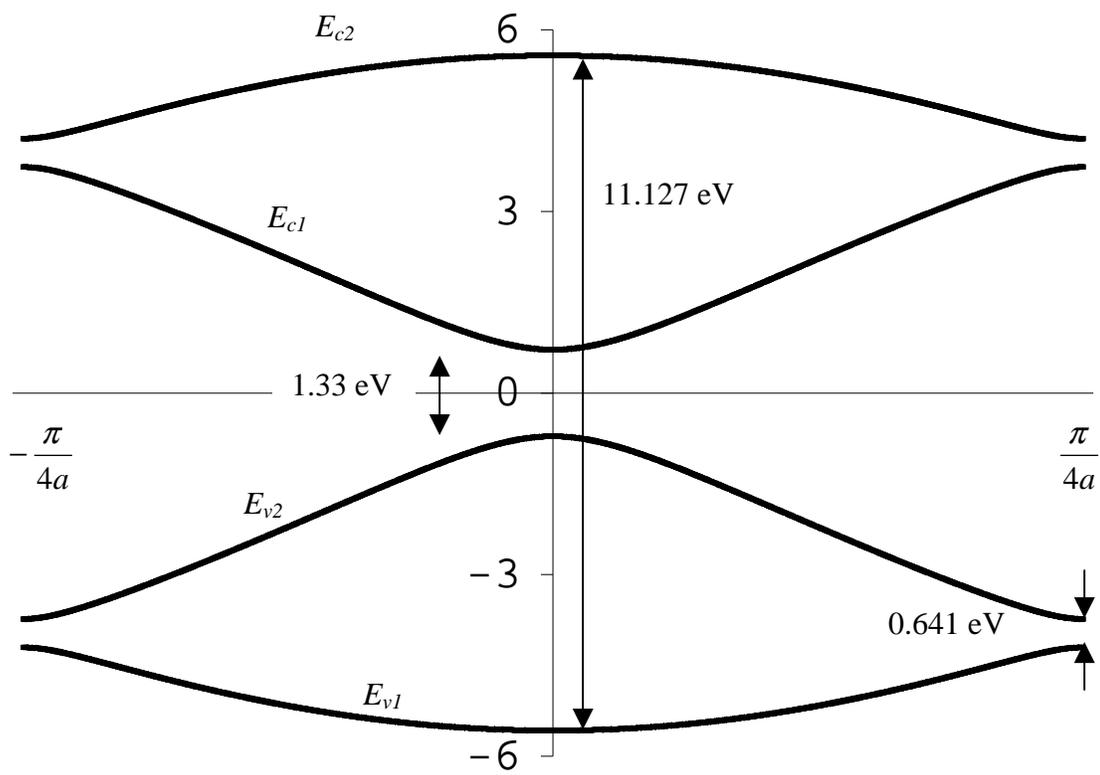
FIG. 4.{\bf} Schematic energy band structure of polydiacetylene in
the reduced Brillouin zone for the acetylene structure. E'$_{g}$
is the energy gap within the valence bands. E$_{g}$ is the energy
gap between the upper valence band and the conduction band.
Numerical estimates are from the parameters of Table 2.0 (see
text).

\bigskip

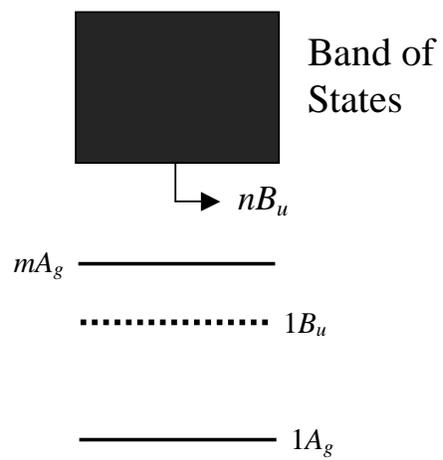
FIG. 5.{\bf}  The four essential states.

\bigskip

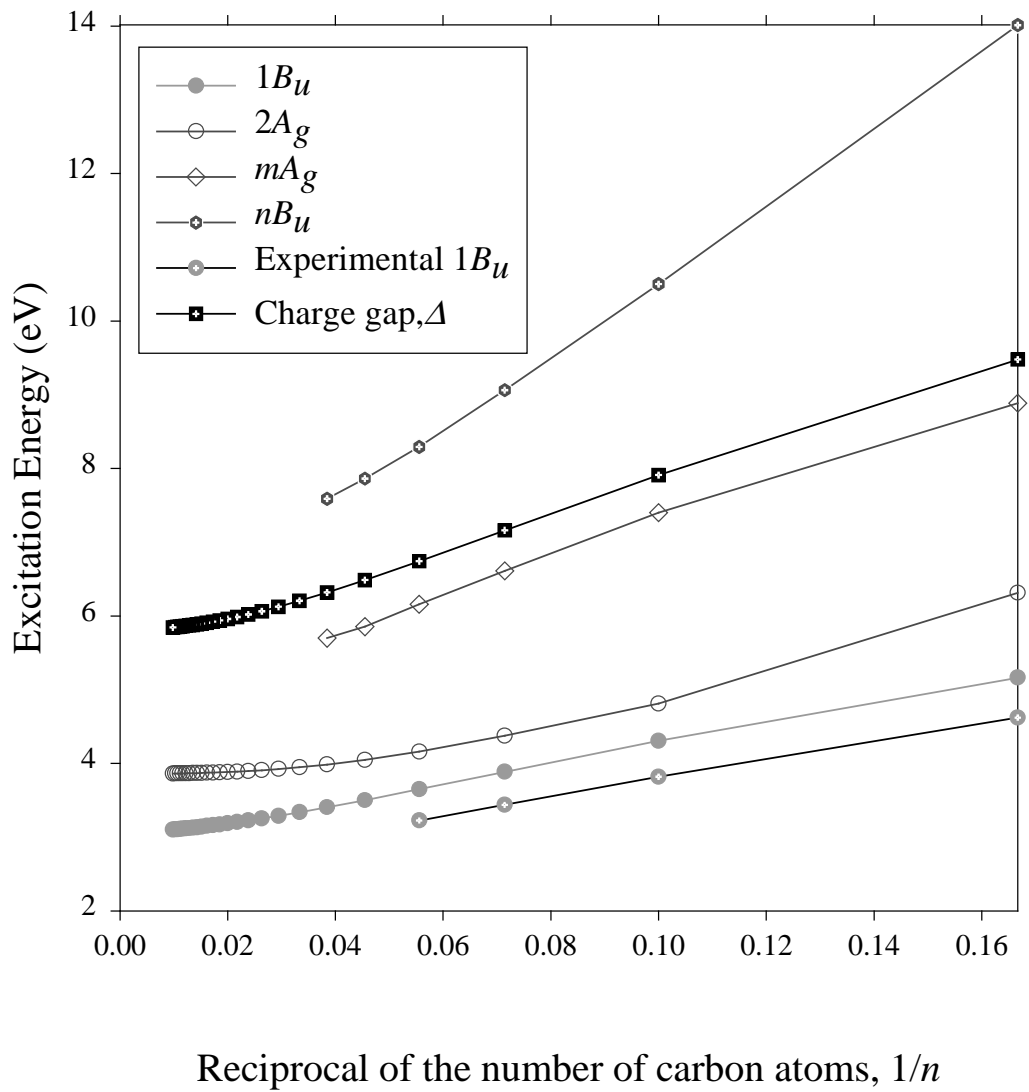
FIG. 6. Excitation energies for the 1{\it B}$_{u}$, 2{\it
A}$_{g}$, {\it mA}$_{g}$ and {\it nB}$_{u} $states for the PPP
model as function of the inverse of the number of carbon atoms,
1/{\it n}. The 1{\it B}$_{u}${\it }experimental excitation
energies are included for comparison.

\bigskip

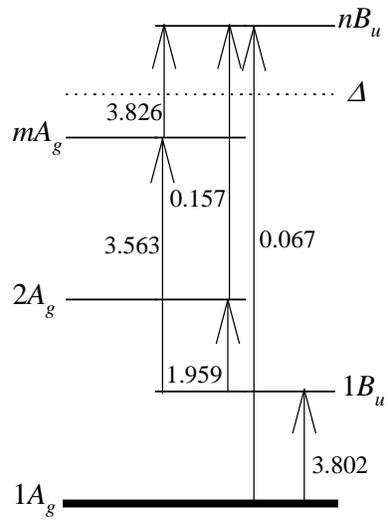
FIG. 7. The states contributing to the non-linear properties of
PDA and the important one-photon transitions between them. The
dipole moments are shown for a chain of 26 atoms.

\bigskip

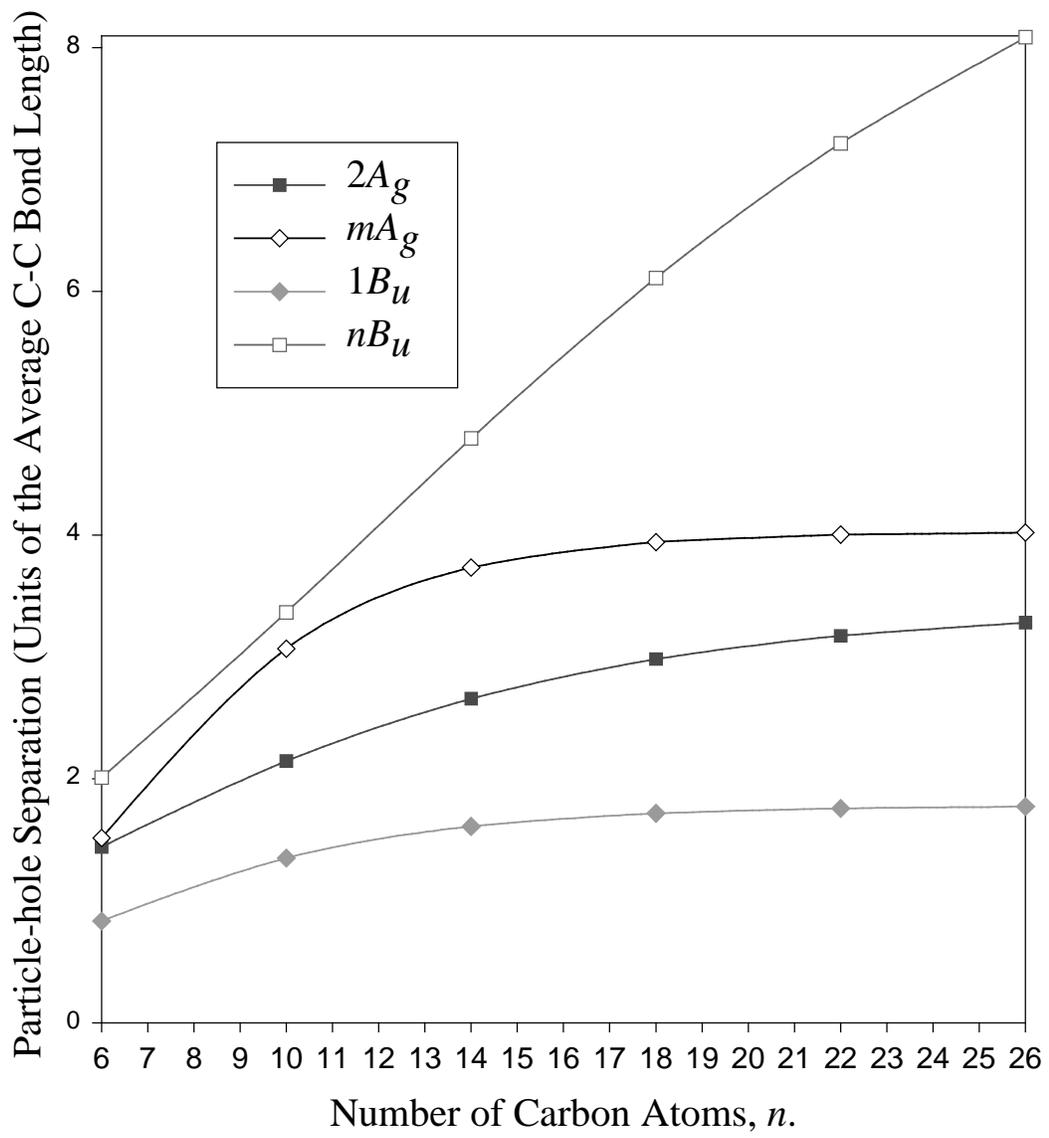
FIG. 8. Particle-hole separation for the 1{\it B}$_{u,}${\it}
2{\it A}$_{g}$, {\it mA}$_{g}$ and {\it nB}$_{u}${\it} states in
units of the average C-C distance as a function of the number of
the number of carbon atoms, {\it n}.

\end{document}